\documentstyle{elsart}

\newcommand{\beq}{\begin{equation}}
\newcommand{\eeq}{\end{equation}}
\newcommand{\bea}{\begin{eqnarray}}
\newcommand{\eea}{\end{eqnarray}}
\newcommand{\non}{\nonumber}
\newcommand{\rf}[1]{(\ref{#1})}

\begin{document}
\begin{frontmatter}
\title{Quantum Computers and Unstructured Search: \\
Finding and Counting Items with an Arbitrarily Entangled Initial State}
\author{A. Carlini and A. Hosoya}
\address{Department of Physics, Tokyo Institute of Technology,
Oh-Okayama, Meguro-ku, Tokyo 152, Japan\\
e-mail: carlini@th.phys.titech.ac.jp ~~;~~ ahosoya@th.phys.titech.ac.jp}


\begin{abstract}
Grover's quantum algorithm for an unstructured search problem and
the Count algorithm by Brassard et al. are generalized to
the case when the initial state is arbitrarily and maximally entangled.
This ansatz might be relevant with quantum subroutines, when the computational
qubits and the environment are coupled,
and in general when the control over the quantum system is partial.


\end{abstract}



\end{frontmatter}



\section{Introduction}

In the recent years quite a significant progress has been made
in the theory of quantum computation [1-3], both at the theoretical
and experimental level [4-5].
In the quest for quantum algorithms, in particular, after the
discovery by Shor \cite{shor} of an algorithm for factoring  integers
(which achieves an exponential speed up compared to the best classical
algorithm currently known),
one of the main successes has been Grover's algorithm for
the unstructured database search \cite{grover1}.
Grover considered the problem to find a 'good' file,
represented as the state $|g>$, out of $N$ files $|a>~~;~~a=0 \ldots N-1$.
The algorithm starts with the preparation of a flat superposition of all
states $|a>$, i.e.\footnote{Assuming, without loss of generality, that $N$
is a power of two.}

\beq
|\psi_0>\equiv {\mbox{\boldmath $W$}}|0>\equiv
{1 \over \sqrt{N}}\sum_{a=0}^{N-1}|a>,
\label{1}
\eeq
where ${\mbox{\boldmath $W$}}$ is the Walsh-Hadamard transform,
and assumes that there is an oracle which evaluates the function $H(a)$, s.t.
$H(g)=1$ for the 'good' state $|g>$, and $H(b)=0$ for the 'bad'
states $|b>$ (i.e., the remaining states in the set of all the $a$'s).
The unitary transformation for the 'search' of $|g>$ is usually defined
[7-8] in terms of the operator
${\mbox{\boldmath $G$}}_H\equiv
-{\mbox{\boldmath $W$}}{\mbox{\boldmath $S$}}_0{\mbox{\boldmath $W$}}
{\mbox{\boldmath $S$}}_H$,
with the inversion
operators ${\mbox{\boldmath $S$}}_0\equiv {\bf I}-2|0><0|$
and ${\mbox{\boldmath $S$}}_H\equiv {\bf I}-2\sum_{g}
|g><g|$.\footnote{In fact ${\mbox{\boldmath $S$}}_H$
can be easily implemented as an $|a>$-'controlled'
unitary transformation ${\mbox{\boldmath $U$}}_H$ by tensoring $|\psi_0>$ with
an extra ancilla qubit $|e>\equiv [|0>-|1>]/\sqrt{2}$,
such that ${\mbox{\boldmath $U$}}_H|a>|e>\equiv
|a>|e+ H(a)\bmod 2>$, thus obtaining
${\mbox{\boldmath $U$}}_H|\psi_0>\rightarrow
{1 \over \sqrt{N}}\sum_{a=0}^{N-1}|a>(-1)^{H(a)}|e>$.
Furthermore, the `inversion about the average' operator
can also be compactly written
as ${\mbox{\boldmath $U$}}_{\psi_0}
\equiv {\mbox{\boldmath $W$}}{\mbox{\boldmath $S$}}_0
{\mbox{\boldmath $W$}}={\bf I}-2|\psi_0><\psi_0|$.}
Iterating ${\mbox{\boldmath $G$}}_H$ for $n\simeq O[\sqrt{N}]$
times on \rf{1} then produces a state whose amplitude
is peaked around the searched item $|g>$.
Classically, it would take of the order of $O[N]$ steps on the average
to find the same element $g$, so that  Grover's quantum method achieves
a square root speed up compared to its classical analogue.
Subsequently, Grover's algorithm has been extended to the case when
there are $t$ 'good' items $|g>$ to be searched
or when the number of 'good' items is not known in advance [8-11].
The number of steps required in these cases is of the order of $O[\sqrt{N/t}]$,
again a square root improvement with respect to the classical
algorithms.
Then, Brassard et al. \cite{brassard}
combined Grover's operator and Shor's quantum Fourier transform
in an algorithm that counts the number of 'good' items present
in the flat superposition \rf{1} with an exponential precision
and a success probability exponentially close to one.
It has then been shown that Grover's algorithm is optimal [13-14],
that a single (complex) oracle query in Grover's algorithm might
suffice in certain cases for finding the 'good' states [15-18],
the algorithm has been exploited in $NP$-structured search problems
[19-20], its Hamiltonian formulation \cite{farhi1} and robustness
discussed [22].
Although Grover's algorithm was originally devised assuming that the
starting state is to be prepared in the {\it flat superposition}
form \rf{1}, subsequent work [8, 11, 23] showed that this condition can
actually be relaxed and that one can also work with the general
initial pure state $|\bar\psi_0>\equiv {\mbox{\boldmath $U$}}|0>$,
i.e. replace ${\mbox{\boldmath $W$}}$ by an arbitrary unitary
transformation ${\mbox{\boldmath $U$}}$, and then use the operator
${\mbox{\boldmath $G$}}_H\equiv
-{\mbox{\boldmath $U$}}{\mbox{\boldmath $S$}}_0{\mbox{\boldmath $U$}}^{-1}
{\mbox{\boldmath $S$}}_H$.
It was also shown \cite{jozsa1} that ${\mbox{\boldmath $G$}}_H$ can be
interpreted in terms of a rotation of $|\bar\psi_0>$ towards the vector
representing the `good' states, $|w>\equiv\sum_g|g>/\sqrt{t}$,
which is a product of two reflections
in the two-dimensional space spanned by  $|\bar\psi_0>$ and $|w>$.
An explicit calculation for the case when the
amplitudes of the initial superposition of states are {\it arbitrary and
unknown complex numbers} was then made by Biham et al. \cite{biham}, who
found that one can still express the optimal measurement time and the maximal
probability of success in a closed and exact form which depends only on the
averages and the variances of the initial amplitude distribution
of states.\footnote{Further discussion for the case of an arbitrary
(non entangled) initial
state and/or an arbitrary unitary transform in Grover's search
can be found in Pati (1998) and Gingrich et al. (1999).}

One of the main resources and ingredients of quantum computation lies,
however, not only in the possibility of dealing with arbitrary
complex superpositions of qubits, but also in the massive
exploitation of {\it quantum entanglement} (see, e.g., refs. [28-29]).
In particular, it is not obvious to which extent one can
directly use Grover's algorithm when the states to be searched
are nontrivially coupled with the states of another system
over which we cannot have a complete control, i.e. when
the initial (normalized) superposition is given by\footnote{We take the
normalization conditions
$\sum_{a=0}^{N-1}<a|a>\equiv N$, $<a|a^{\prime}>=\delta_{aa^{\prime}}$,
$\sum_{a=0}^{N-1}<f_a|f_a>\equiv N$ and
$\sum_{g}<f_g|f_g>\equiv N_1$,
with $N_1\leq N$ and, in general, $<f_a|f_b>\not = \delta_{ab}$.}

\beq
|\psi>\equiv {1\over \sqrt{N}}\sum_{a=0}^{N-1}|a>|f_a>,
\label{3}
\eeq
where $f_a$ is an arbitrary mapping (not necessarily one to one).
In fact, in Grover's algorithm the application of any unitary
transformation acting on the computational
states $|a>$ would automatically affect {\it also
and nontrivially} the $|f_a>$'s, by producing a complicated mixing of the
original entanglements.
An interesting case may arise, for instance, when the system $|f_a>$
represents the environment.
Of course, one might trace out (measure) the `environment' and
simply work with the $|a>$ subsystem, which after the measurement of the
$|f_a>$ subsystem will collapse into a generic mixed state, i.e.
an incoherent superposition of sectors of pure states (each characterized by
a different value of $f_a$), and then simply study the dynamics in each
of these sectors using the methods described, e.g., in ref. \cite{biham}.
The problem with this approach is that sectors with
different $f_a$'s will generally have different numbers of `good' states,
and thus require different computational time to find the `good' states
(in some cases failing, for the sectors where there are no `good' states).
One can give an upper
bound on the success probability of the algorithm, which cannot exceed
the expectation value of the operator ${\bf I}_a\sum_{g, g^{\prime}}
|f_g><f_{g^{\prime}}|$, however, for generic situations
in principle one could do much worse
and there is no explicit scheme telling what one should expect.
Alternatively, one might consider the setting in which Alice prepares
the state \rf{3}, where the function $f_a$ is complicated and
requires a lengthy computation, and sends it to Bob.
Bob then wants to access the information contained in the $|f_g>$'s
associated with certain marked states $|g>$, under the promise by Alice
that there are $t$ `good' $|g>$'s, but without having an a priori
`local' knowledge about the $f_a$'s or how to compute them (except, possibly,
the knowledge about certain `global' properties such as the first order
momenta of the $|f_a>$'s), and the simplest way to do that
is to enhance the amplitude of the $|g>$'s using Grover's algorithm
first and then read the associated $|f_a>$.
Finally, in number theory one might want
to count (for example, when testing the Prime Number theorem, see
ref. \cite{hardy}) the number of primes $k$ less than a given integer
$N$, and then need the quantum (entangled) superposition
$\sum_{k=0}^{N-1}|k>|f(k)>$, where $f$ is a `primality flag'
for $k$ (i.e., $f=1$ for $k$ prime, and $f=0$ for $k$ composite), which
itself maybe the result of a previous lengthy quantum subroutine
and which we want to reuse in further computations.

The aim of the present paper is thus to generalize Grover's
methods to the case when the initial
state is given by the entangled superposition \rf{3}.

\section{Finding Good States}

The problem is to find the `good' states $|g>$, promised to be in number $t$,
from the initial normalized entangled superposition \rf{3}.
By defining the remaining or `bad' states as $|b>$, in number $N-t$, and
applying Grover's unitary transformation
${\mbox{\boldmath $G$}}_H\equiv -{\mbox{\boldmath $W$}}{\mbox{\boldmath $S$}}_0
{\mbox{\boldmath $W$}}{\mbox{\boldmath $S$}}_H$
on the state $|\psi>$ of eq. \rf{3},
it is easy to show (by induction) that the $n$-th iteration of
${{\mbox{\boldmath $G$}}_H}$ on $|\psi>$ produces the state

\bea
{{\mbox{\boldmath $G$}}^n_H}|\psi>&=&{1\over \sqrt{N}}
\biggl \{ [\sum_{g}|g>|f_g>]+(-1)^n [\sum_{b}|b>|f_b>]
\non \\
&-&{2\over N}\left [(\sum_{g}|g>)|X^{(n)}>+(\sum_{b}|b>)|Y^{(n)}> \right]
\biggr \}.
\label{6}
\eea
Adopting a compact matrix notation, i.e. by substituting for
$|X^{(n)}>\rightarrow X_n$ and $|Y^{(n)}>\rightarrow Y_n$,
writing $\overrightarrow{Z}_n\equiv (X_n, Y_n )$ and defining the matrix
${\mbox{\boldmath $M$}}\equiv \cos 2\theta ({\mbox{\boldmath $I$}}+
{\mbox{\boldmath $\sigma$}}_x)+i{\mbox{\boldmath $\sigma$}}_y $
(with the angle $\sin^2 \theta \equiv t/N$),
the states $|X^{(n)}>$ and $|Y^{(n)}>$
can be seen to satisfy the recurrence relations

\beq
\overrightarrow{Z}_n={\mbox{\boldmath $M$}}\overrightarrow{Z}_{n-1}+
\overrightarrow{C}_n,
\label{8}
\eeq
which depend only on the number of `good' items via the angle $\theta$ and
the following initial `average' states

\beq
|{\bar G}^{(0)}>\equiv {\sum_{g}|f_g>\over t}~~~;~~~
|{\bar B}^{(0)}>\equiv {\sum_{b}|f_b>\over N-t}
\label{4}
\eeq
via the quantities
$|C^{(n)}>\equiv t|{\bar G}^{(0)}>+(-1)^n(N-t)|{\bar B}^{(0)}>$, with
the substitution $|C^{(n)}>\rightarrow C_n$ and writing
$\overrightarrow{C}_n\equiv C_n(1, 1 )$
(subject to the the initial condition $X_1=Y_1=C_1$).
Eq. \rf{8} can then be solved using standard techniques and, substituting
back in eq. \rf{6} we finally get for the $n$-th iteration
of ${{\mbox{\boldmath $G$}}_H}$ on the entangled state $|\psi>$

\bea
{{\mbox{\boldmath $G$}}^n_H}|\psi>&=&{1\over \sqrt{N}}
\biggl \{ \sum_{g}|g>\biggl [|f_g>-{\sin 2n \theta\over \sin 2\theta}
\biggl (\tan n\theta \sin 2\theta |{\bar G}^{(0)}>
\non \\
&-&2\cos^2\theta|{ \bar B}^{(0)}>\biggr )\biggr ]
+\sum_{b}|b>\biggl [(-1)^n|f_b>
\non \\
&-&{\sin 2n \theta\over \sin 2\theta}
\biggl (2\sin^2\theta |{\bar G}^{(0)}>
+(-1)^n\sin 2\theta(\tan n\theta)^{(-1)^n}|{\bar B}^{(0)}>
\biggr )\biggr ]\biggr \}
\non \\
&\equiv &{1\over \sqrt{N}}
[\sum_{g}|g>|f^{(n)}_g>+\sum_{b}|b>|f^{(n)}_b>].
\label{12}
\eea
As we can already note from expression \rf{12}, similarly to the case of the
original Grover's
algorithm acting on an initial flat superposition of states,
also in the presence of entanglements ${{\mbox{\boldmath $G$}}^n_H}|\psi>$ is
periodic in $n$ with period $\pi/\theta$, and one can anticipate that
a Fourier analyis
can still be performed in order to find an estimate of $\theta$ (see next
section).

It is then possible to show, following methods similar to those used in
ref. \cite{biham} and after some elementary algebra, that the `variances'
of the distribution of the amplitudes of the initial entangled state are
constants of the motion, i.e.

\bea
\sigma^{2 (n)}_{G}&\equiv &{\sum_{g}\parallel
(|f^{(n)}_g>-|{\bar G}^{(n)}>)\parallel^2\over t}=\sigma^{2 (0)}_{G}
\equiv \sigma^2_G
\non \\
\sigma^{2 (n)}_{B}&\equiv &{\sum_{b}\parallel
(|f^{(n)}_b>-|{\bar B}^{(n)}>)\parallel^2\over N-t}=\sigma^{2 (0)}_{B}
\equiv \sigma^2_B,
\label{17}
\eea
where $\parallel |x>\parallel^2 \equiv <x|x>$, and
$|{\bar G}^{(n)}>\equiv [\sum_{g}|f^{(n)}_g>]/t$ and
$|{\bar B}^{(n)}>\equiv [\sum_{b}|f^{(n)}_b>]/(N-t)$ are the `averages'
at time $n$.

Using this fact, one can calculate
the probability of picking up a 'good' item after $n$ iterations
of ${{\mbox{\boldmath $G$}}_H}$ over the initial entangled state
$|\psi>$, defined by $P(n)\equiv \sum_{g}<f^{(n)}_g
|f^{(n)}_g>$, as

\bea
P(n)&\equiv &P_{AV}-\Delta P\cos 2(2n\theta -\phi_R)e^{-2\phi_I}
\non \\
P_{AV}&\equiv &1-\Delta P -N\sigma_B^2\cos^2\theta
\non \\
\Delta P&\equiv &{N\over 2}\cos^2\theta\biggl [
<{\bar B}^{(0)}|{\bar B}^{(0)}>+
\tan^2\theta<{\bar G}^{(0)}|{\bar G}^{(0)}>\biggr ],
\label{24}
\eea
where the complex angle $\phi\equiv \phi_R+i\phi_I$ is defined by the formula

\beq
\exp[2i\phi]\equiv {2<F^{(0)}_+|F^{(0)}_->\over [<F^{(0)}_+|F^{(0)}_+>
+<F^{(0)}_-|F^{(0)}_->]},
\label{21}
\eeq

with $|F^{(0)}_{\pm}>\equiv|{\bar B}^{(0)}>\pm i\tan \theta
|{\bar G}^{(0)}>$.
As it can be easily seen from eqs. \rf{24} and \rf{21}, the probability
$P(n)$ only depends on the first order momenta of the distribution of
the amplitudes of the initial entangled state, i.e. upon
the initial `averages' $<{\bar G}^{(0)}|{\bar G}^{(0)}>$,
$<{\bar B}^{(0)}|{\bar B}^{(0)}>$ and
the initial `variance' $\sigma_B^2$, and on the number of `good' items $t$.
This is similar to the case of a non entangled initial state with arbitrary
amplitudes \cite{biham}.
The probability $P(n)$ is maximized, $P_{MAX}=P_{AV}
+\Delta P~e^{-2\phi_I}$, at the times

\beq
n_j= [\pi(2j+1)/2+\phi_R]/2\theta ~~~~;~~~~j\in {\cal Z},
\label{24b}
\eeq
and the minimum number of iterations required to achieve the maximal
probability of success $P_{MAX}$ is $n_0$.
In particular, for the case of $\theta \simeq \sqrt{t/N}\ll 1$, we have that
$n_0\simeq O(\sqrt{N/t})$, which is the same as in Grover's original
algorithm.
Moreover, one can almost be certain to find a 'good' item
$|g>$ only provided that (for $\theta \ll 1$) the initial `variance'
of the amplitudes of the `bad' states is small enough,
i.e., if $N\sigma_B^2\equiv \epsilon \ll 1$ then we have that
$P_{MAX}\simeq 1 -\epsilon \simeq O(1)$, independently of the values of
the initial `averages' $<{\bar G}^{(0)}|{\bar G}^{(0)}>$ and
$<{\bar B}^{(0)}|{\bar B}^{(0)}>$.
\footnote{The probability of finding the `good' states does not change
with the number of trials $n$
(besides the trivial cases of $\theta=\pi/2$, $t=N$, $\theta=t=0$ or, e.g.,
when the function $f_a$ is one to one,
since then the reduced density matrix
$\rho_a\equiv Tr_f|\psi><\psi|=\sum_a|a><a|$, on which Grover's
operator acts, is totally mixed, i.e. $\rho_a=I$,
and so invariant under any unitary transformation)
when $|<F^{(0)}_+|F^{(0)}_->|=0$,
since then we get $P(n)=P_{AV}=1-N\sigma_B^2\cos^2\theta =const $.
The probability of finding the `good' states can be small, i.e.
$P(n)\simeq \delta \ll 1$, if the initial variance $\sigma_B^2\simeq
(1-\delta)/(N-t)$ ($\sigma_B^2\approx N^{-1}$ for $t\ll N$).
Finally, the algorithm completely fails only when one of the previous
conditions holds
and there is a fine tuning $N\sigma_B^2\cos^2\theta =1$,
for which $P(n)=const=0$.}
A more detailed analysis of the possible outcomes when the `averages'
and the `variances' of the amplitudes of the initial entangled state
$|\psi>$ are arbitrary and unknown (e.g., best average number of steps
necessary to find the 'good' states $|g>$, etc..) can be done following the
same lines as in ref. \cite{biham}.
It is straightforward to show that the model of an
initial state with arbitrary complex amplitudes described in ref. \cite{biham}
can be recovered as a subcase of our algorithm
provided one makes the substitutions $|f_g>=k_i|0>$ and
$|f_b>=l_i|0>$ (with $k_i$ and $l_i$ arbitrary complex phases).
This includes, in particular, Grover's original ansatz \cite{grover1}
for the choice $k_i=l_i=1$.

Finally, it is also interesting to note that, for $n=n_j$,
we get $\sqrt{N}{{\mbox{\boldmath $G$}}^{n_j}_H}
|{\psi}>=\sum_g|g>[|f_g>-(1+(-1)^j\sin\phi_R)
|{\bar G}^{(0)}>+(-1)^j\cos\phi_R\cot\theta
|{\bar B}^{(0)}>]$.
Of course, unitarity prevents one to naively get only the exact contribution
from the initial `unperturbed' entangled states $|f_g>$
in ${{\mbox{\boldmath $G$}}^{n_j}_H}|{\psi}>$.
However, as some elementary algebra can show, it is still possible, for
instance in the case of a large enough number of `good' items $g$,
i.e. for $t/N\leq O(1)$, to make the amplitude contribution coming
from the other entangled states $|{\bar G}^{(0)}>$ and
$|{\bar B}^{(0)}>$ relatively small compared to that of
$|f_g>$, if $j$ is even and provided that
$<{\bar G}^{(0)}|{\bar G}^{(0)}>\simeq <{\bar B}^{(0)}|{\bar B}^{(0)}>$.

\section{Conclusions}

We have shown how to generalize one of the most useful algorithms
discovered in quantum computing so far, i.e. Grover's algorithm
for the search into an unstructured database, in the most general
case when this algorithm is to be applied on a {\it generic initial state in
an unknown, arbitrary and maximally entangled superposition of qubits}.
This situation might arise, for example, when the computational qubits
are nontrivially entangled with the environment, when one wants to search or
count marked items in subroutines part of larger quantum networks,
or more in general when part of the quantum system is not accessible.
In particular, we have seen that even in this general case, the
dynamics of the quantum entangled system is periodic, with the
period depending on the number of `good' items, and fixed by the
first order momenta of the amplitudes of the initial state.
Furthermore, the search algorithm still generically needs $O(\sqrt{N/t})$
iterations to sort one of the `good' items, and
the maximum probability to obtain such  `good' items can be made close to
one, provided that the initial `variance' of the amplitude distribution of the
`bad' states is small enough.
In the appendix we
also generalized the COUNT algorithm and showed that one can preserve
a good success probability and a high accuracy in determining
the number of `good' items even if the initial state is entangled,
provided that some conditions are satisfied by the `averages'
of the amplitude distribution of the initial state (for example,
for the choice
$<{\bar G}^{(0)}|{\bar G}^{(0)}>~\geq ~<{\bar B}^{(0)}|{\bar B}^{(0)}>~>
\pi^2/(8\sqrt{2})$).
These results were not obvious a priori and constitute a non trivial
feature for quantum computational states with entanglement.

\vspace{33pt}

\noindent {\Large \bf Acknowledgements}

\bigskip
A.H.'s research was partially supported by the Ministry of Education,
Science, Sports and Culture of Japan, under grant n. 09640341.
A.C.'s research was supported by the EU under the Science and Technology
Fellowship Programme in Japan, grant n. ERBIC17CT970007; he also thanks
the cosmology group at Tokyo Institute of Technology for the kind hospitality
during this work.

\appendix
\section{Counting Good States}

The algorithm COUNT, introduced by Brassard et al. \cite{brassard}
for the case of an initial flat superposition of states,
essentially exploits Grover's unitary operation
${{\mbox{\boldmath $G$}}_H}$,
already discussed in the previous
section, and Shor's Fourier operation ${\mbox{\boldmath $F$}}$
for extracting the periodicity of a quantum state, defined
as\footnote{We assume, without loss of generality, that $k$ is a power
of 2.}

\beq
{\mbox{\boldmath $F$}}|a>\equiv
{1\over \sqrt{k}}\sum_{c=0}^{k-1}e^{2i\pi ac/k}|c>.
\label{26}
\eeq
Then, the COUNT algorithm can be summarized by the following
sequence of operations:
~~1) $({\mbox{\boldmath $W$}}|0>)({\mbox{\boldmath $W$}}|0>)
=\sum_{m=0}^{P-1}|m>\sum_{a=0}^{N-1}|a>$
~~2) $\rightarrow ({\mbox{\boldmath $F$}}\otimes {\mbox{\boldmath $I$}})
[\sum_{m=0}^{P-1}|m>{\mbox{\boldmath $G$}}_H^m(\sum_{a=0}^{N-1}|a>)]$
~~3) $\rightarrow \mbox{measure} ~~|m>$.

The main idea at the core of the algorithm is that, since the amplitude
of the set of the good states $|g>$ after $m$ iterations of
${\mbox{\boldmath $G$}}_H$ on $|a>$ is a periodic function of $m$,
the estimate of such a
period by use of the Fourier analysis and the measurement of the
ancilla qubit $|m>$ will give information on the size $t$ of this set,
on which the period itself depends.
The parameter $P$
determines both the precision of the estimate $t$ and the computational
complexity of the COUNT algorithm (which requires $P$ iterations of
${\mbox{\boldmath $G$}}_H$ \cite{brassard}).

Let us now discuss how one can use the previous results, obtained
for the generalized Grover's algorithm when the initial state is
arbitrarily entangled, to the case of the COUNT algorithm.
We start by tensoring the state $|{\psi}>$
with $\log P$ ancilla qubits\footnote{Without loss of generality,
we assume that $P$ is a power of $2$.} set to $|0>$, act on these qubits
with ${\mbox{\boldmath $W$}}$, obtaining $|{\psi}_1>\equiv \sum_{m=0}^{P-1}
|m>|{\psi}>/\sqrt{P}$, with an $|m>$-'controlled' Grover operation
${{\mbox{\boldmath $G$}}^m_H}$ on the state $|\psi>$ and with
${\mbox{\boldmath $F$}}$ on $|m>$, thus getting

\beq
|{\psi}_f>\equiv {\sum_{m, n=0}^{P-1}e^{2\pi i mn/P}|n>\over \sqrt{P}}
{{\mbox{\boldmath $G$}}^m_H}|{\psi}>.
\label{28}
\eeq
As in the standard COUNT algorithm, requiring that
the time needed to compute the repeated Grover operations
${{\mbox{\boldmath $G$}}^m_H}$ is polynomial in $\log k$,
leads to the choice $P\simeq O[poly (\log k)]$ in eq. \rf{28}.
Explicitly
summing over $n$ in eq. \rf{28}, after some elementary algebra we get

\bea
|{\psi}_f>&\equiv & {1\over \sqrt{N}}\biggl [|0>|A>+|P/2>|B>
+{1\over 2N}\sum_{m=0}^{P-1}|m>(\phi^+_{m}s^+_{m}|C_+>
\non \\
&+&\phi^-_{m}s^-_{m}|C_->)\biggr ],
\label{29}
\eea
where we have introduced the following quantities

\beq
s^{\pm}_m\equiv {\sin\pi (m\pm f)\over P\sin[\pi(m\pm f)/P]}~~;~~
\phi^{\pm}_m\equiv e^{i\pi (m\pm f)(1-1/P)}~~;~~f \equiv {P\theta\over \pi},
\label{30}
\eeq
with $0\leq f\leq P/2$, and $|A>, |B>$ and  $|C_{\pm}>$ are
certain mutually orthogonal states.

At this point one can rewrite formula \rf{29},
similarly to how it is explained in ref. \cite{brassard},
in the general case when $f$ is not an integer,
distinguishing three possible cases.
In particular, in the most general case in which $1<f<P/2-1$ we have

\bea
|{\psi}_f>&=&|f^->|a_1>+|P-f^->|b_1>+|f^+>|c_1>
\non \\
&+&|P-f^+>|d_1>+|R_1>,
\label{37}
\eea
where $|R_1>$ is an 'error' term including all the other states in
$|{\psi}_f>$ not containing the ancilla qubits $|f^{\pm}>$ and
$|P-f^{\pm}>$ (with $f^-\equiv [f]+\delta f$, $f^+\equiv f^- +1$
and $0<\delta f<1$).
After some easy algebra, one can show that the total probability
amplitude in the first four terms
is given by

\beq
W_1=[\sin^2\theta <{\bar G}^{(0)}|{\bar G}^{(0)}>
+\cos^2\theta <{\bar B}^{(0)}|{\bar B}^{(0)}>]\Sigma_1,
\label{38}
\eeq
with $\Sigma_1\equiv ({s^+_{f^+}})^2+({s^+_{f^-}})^2+({s^+_{P-f^+}})^2
+({s^+_{P-f^-}})^2$, and it can be shown that $8/\pi^2<\Sigma_1\leq 1$.
Similar calculations can be done for the cases when $0<f<1$, when the
probability of obtaining any of the states $|0>$, $|1>$ or $|P-1>$ is
given by
$W_2=\{N_1 +N[(\Sigma_2-1)\sin^2\theta<{\bar G}^{(0)}|{\bar G}^{(0)}>
+\Sigma_2\cos^2\theta <{\bar B}^{(0)}|{\bar B}^{(0)}>]\}/N$
(with
$\Sigma_2\equiv ({s^+_0})^2+({s^+_1})^2+({s^+_{P-1}})^2$
and $8/\pi^2<\Sigma_1\leq 1$), or when $P/2-1<f<P/2$, for which
the probability to get any of the states $|P/2>$ or $|P/2\pm 1>$ reads
$W_3=1-\{N_1 - N[\Sigma_3\sin^2\theta<{\bar G}^{(0)}|{\bar G}^{(0)}>
+(\Sigma_3-1)\cos^2\theta <{\bar B}^{(0)}|{\bar B}^{(0)}>]\}/N$
(with $\Sigma_3\equiv ({s^+_{P/2}})^2+({s^+_{P/2-1}})^2+({s^+_{P/2+1}})^2$
and $8/\pi^2<\Sigma_3\leq 1$).

The final step of the COUNT algorithm consists in measuring the
first ancilla qubit in the state $|{\psi}_f>$.
As explained in ref. \cite{brassard}, if $W_i\geq 1/2$ (for $i=1, 2$ or $3$)
then with high probability\footnote{By repeating the whole algorithm many
times and using the majority rule, Brassard et al. (1998),
and eventually using
$R$ ancilla qubits $|m_1>...|m_R>$ and acting with a `$|m_1>...|m_R>$
-controlled' ${{\mbox{\boldmath $G$}}^m_H}$
operation on the state $|{\psi}>$, Carlini et al. (1999).}
one can still be able to
find one of the ancilla qubits $|f_{\pm}>$ or $|P-f_{\pm}>$, $|0>, |1>$ or
$|P-1>$, $|P/2\pm 1>$ or $|P/2>$, respectively, for the three cases,
and, therefore, evaluate the number $t$ of 'good' states
from $\sin\theta =\sqrt{t/N}$ and eq. \rf{30}.
For example, in the general case $1<f<P/2-1$, the condition that $W_1>1/2$
is satisfied, e.g., for the choice of the initial `averages'
$<{\bar G}^{(0)}|{\bar G}^{(0)}>~\geq ~<{\bar B}^{(0)}|{\bar B}^{(0)}>~>
\pi^2/(8\sqrt{2})$.
Finally, although in general $f$ is not an integer and the measured
${\tilde f}$ will not match exactly the true value of $f$ but give
the approximate estimate ${\tilde t}\equiv N\sin^2{\tilde \theta}({\tilde f})$,
the error over $t$ for an entangled initial state will be
$|{\tilde t}-t|\leq\pi N\left [\pi/P + 2\sqrt{t/N}\right ]/P$,
i.e. the same as in ref. \cite{brassard}.

\end{document}